\RequirePackage{fix-cm}
\documentclass[smallextended]{svjour3}       % onecolumn (second format)
\smartqed  % flush right qed marks, e.g. at end of proof
\usepackage{graphicx}
\usepackage{mathptmx}      % use Times fonts if available on your TeX system
\usepackage{setspace}
\usepackage{amsmath}
\usepackage{amssymb}
\usepackage{graphicx}
\usepackage{mathrsfs}
\usepackage{mathrsfs}
\usepackage{bm}
\usepackage{wasysym}
\usepackage{lineno}
\usepackage[round]{natbib}
\bibpunct{(}{)}{;}{a}{}{;}

\journalname{Theoretical Ecology}

\begin{document}

\linenumbers 
\modulolinenumbers[2]

%\title{Estimating the degree of compensation from fluctuations in fish biomass, or why the Beverton Holt function cannot be measured in nature}
%\title{Relating the degree of compensation to functional elasticities and why the Beverton-Holt cannot be measured in nature}
%\title{Using functional elasticities to determine the compensatory dynamics of fish recruitment, and why the Beverton-Holt Stock Recruitment Relationship cannot exist in nature}
%\title{Compensatory dynamics, functional elasticities, and why the Beverton-Holt Stock Recruitment Relationship cannot exist in nature}
%\title{Relating the compensatory dynamics of fish recruitment to functional elasticities: a generalized modeling approach}
\title{Compensatory dynamics of fish recruitment illuminated by functional elasticities}

\titlerunning{Relating the compensatory dynamics of fish recruitment to functional elasticities}        % if too long for running head

\author{Justin D. Yeakel \and
Marc Mangel
}
\institute{J.D. Yeakel \at 
Center for Stock Assessment Research \& \\
Department of Ecology and Evolutionary Biology,\\ 
University of California Santa Cruz, Santa Cruz, CA 95064, USA.\\
\email{jdyeakel@gmail.com}\\
Tel:+1.831.706.4215
\and
M. Mangel \at
Center for Stock Assessment  Research \& \\
Department of Applied Mathematics and Statistics,\\ 
University of California Santa Cruz, Santa Cruz, CA 95064, \\
USA and
Department of Biology, University of Bergen, Bergen 5020, Norway\\
\email{msmangel@ucsc.edu}\\
}

\authorrunning{J.D. Yeakel \& M. Mangel} % if too long for running head

\date{Received: date / Accepted: date}

\maketitle

\begin{abstract}
%Time to write an abstract, lad
Models of Stock Recruitment Relationships (SRRs) are often used to predict fish population dynamics.
Commonly used SRRs include the Ricker, Beverton-Holt, and Cushing functional forms, which differ primarily by the degree of  density dependent effects (compensation).
The degree of compensation determines whether recruitment respectively decreases, saturates, or increases at high levels of spawning stock biomass.
In 1982 J.G. Shepherd united these dynamics into a single model, where the degree of compensation is determined by a single parameter, however the difficulty in relating this parameter to biological data has limited its usefulness.
Here we use a generalized modeling framework to show that the degree of compensation can be related directly to the functional elasticity of growth, which is a general quantity that measures the change in recruitment relative to a change in biomass, irrespective of the specific SRR.
We show that the elasticity of growth can be calculated from short-term fluctuations in fish biomass, is robust to observation error, and can be used to determine general attributes of the SRR in both continuous time production models, as well as discrete time age-structured models.
This framework may be particularly useful if fisheries time-series data are limited, and not conducive to determining functional relationships using traditional methods of statistical best-fit. %Our method of using elasticities to determine the general form of the SRR
%Finally, we show that the Beverton-Holt SRR is qualitatively different than either Ricker-like or Cushing-like SRRs, and that it is impossible for a  Beverton-Holt SRR to exist in  nature.
\end{abstract}


\doublespacing
\section{Introduction}

Recruitment plays a central role in the population dynamics of fish species.
Models of fish recruitment include both density-independent and -dependent effects, controlled by the variables $\alpha$ and $\beta$, respectively.
That is, when density-dependent effects are negligible, recruitment is generally modeled as $S(B) = \alpha B$, where $S(B)$ is the level of recruitment when $B$ is spawning stock biomass, and $\alpha$ is the recruitment rate in the absence of density-dependent effects \citep[e.g.][]{Sissenwine:1987td}. %(e.g Sissenwine and Shepherd 1987 -- this is cited in the accepted version of the steepness paper)
We note that recruitment functions are often introduced as $R(B)$, but to prevent confusion later on (where we introduce scaled functions denoted by lowercase letters, such that $r$ could be confused as a growth rate), we avoid the use of $R$ to denote recruitment.  In using spawning biomass, we have followed a relatively standard assumption of fishery science that fecundity is proportional to biomass.

When density-dependent effects are non-negligible, recruitment is anticipated to deviate from this relationship, such that $S(B) = \alpha B F(\beta,B)$, where the function $F$ controls density-dependent effects on recruitment.
Traditional stock recruitment models introduce three general kinds of density-dependent responses to increasing spawning stock biomass:
1) recruitment increases to a maximum and then declines as $B$ increases,
2) recruitment saturates as $B$ increases,
3) as $B$ increases, recruitment continues to increase but at a lower rate than in the absence of density-dependent effects.
These alternative scenarios thus differ in the intensity of density dependence (degree of compensation), which determines to what extent recruitment is altered as a function of spawning stock biomass.

%JUSTIN -- I WONDERED READING IT THIS TIME IF WE SHOULD USE R(B) FOR RECRUITS, SINCE OFTEN S IS USED FOR SPAWNING STOCK BIOMASS.  THEN WE WOULD WRITE 
%dB/dt = R(B)-D(B), ETC.  BUT THEN OUR SCALED VARIABLE IS r WHICH COULD EASILY BE CONFUSED WITH THAT OF THE LOGISTIC (TRUST ME, SOMEBODY COULD DO IT),
%WHAT DO YOU THINK?  I ALSO THINK THAT THIS WOULD BE A GOOD QUESTION TO ASK ALEC.

%The 3 production models
%The degree of compensation is dependent on the evolutionary ecology of fish species. %and is therefore often species-specific.
%The population-level effects of different compensatory scenarios are typically modeled from separate - sometimes unrelated - Stock Recruitment (SR) functions \citep{Mangel:2006wa}.

\citet{Ricker:1954wpa} developed a Stock-Recruitment Relationship (SRR) to introduce declines in recruitment at high levels of spawning stock biomass (Fig. \ref{figcompdist}a), 

%Ricker Equation
\begin{equation}
%{\rm \frac{d}{dt}}B = \alpha B {\rm e}^{-\beta B},
S(B) = \alpha B {\rm e}^{-\beta B}.
\label{eqRick}
\end{equation}

\noindent As spawning stock biomass increases, recruitment increases to the maximum $S(B) = 1/\beta$ and then declines.
The Ricker model is used if there are predatory response lags, when greater stock abundance suppresses juvenile growth, or when cannibalism or nest predation limits recruitment when $B$ is high \citep{Cushing:1988vb}.

\citet{Beverton:1957we} introduced a related two parameter model where

%Beverton-Holt equation
\begin{equation}
S(B) = \frac{\alpha B}{1 + \beta B},
\label{eqBH}
\end{equation}

%\noindent and $\alpha$ and $\beta$ are as defined previously. 
\noindent Recruitment is thus a saturating function of spawning stock biomass, where saturation occurs at $S(B)=\alpha/\beta$ as $B\rightarrow\infty$.
\noindent Here, it is assumed that density-dependent mortality affects recruitment instantaneously \citep[see][]{Mangel:2006tm}, and that recruitment tends asymptotically towards a finite value as $B$ increases \citep{Cushing:1988vb}.
The Beverton-Holt (B-H) relationship is typically used if recruitment is assumed to be limited primarily by food or habitat resources.  Note that when $\beta B$ is small, so that the exponential in Eq. (1) or the denominator in Eq. (2) are Taylor expanded, we obtain $S(B) = \alpha B(1-\beta B)$, thus giving interpretation to the standard logistic model of population growth. 
%for scenarios where recruitment is primarily a function of limitations imposed by food or habitat resources,

In open systems where resources are not locally limiting, \citet{Cushing:1973wg} developed a power-law SRR

%Cushing equation
\begin{equation}
S(B) = \alpha B (\beta B)^{-1/n_c} = \alpha \beta^{-1/n_c} B^{\frac{n_c-1}{n_c}}.
\label{eqCush}
\end{equation}

\noindent Here, the third parameter $n_c$  controls the rate of recruitment increase at high biomass densities, or the degree of compensation.
In this case, if $n_c>1$ recruitment continues to increase with increasing spawning biomass, but at a decreasing rate.
%the slope of the recruitment curve cannot be $\leq 0$, and therefore cannot incorporate declining or saturating recruitment at high biomass densities.
%Although \citept{Shepherd:1982p3428} described Eq. (\ref{eqCush}) as `pathological' (because $S(B)$ has an infinite slope at the origin if $n>1$), it serves to illustrate the third possible compensatory dynamic common among recruitment functions. JUSTIN -- I DO NOT SEE THIS AT ALL. IT WOULD BE TRUE IF n <1. IN ANY CASE, DONT WE WANT n>1 for our argument? ALSO, SHOULD WE USE A DIFFERENT VARIABLE, SO AS NOT TO CONFUSE WITH THE n OF SHEPHERD?

In an attempt to integrate the above relationships into a single function controlled by the degree of compensation, \citet{Shepherd:1982p3428} observed that the behaviors exhibited by the Ricker, Cushing, and Beverton-Holt functions can be united into a single framework with three free parameters

\begin{equation}
S(B) =  \frac{\alpha B}{1 + \beta B^{1/n}}, ~~\mbox{for}~~n > 0.
\label{eqShep}
\end{equation}

\noindent The parameters $\alpha$ and $\beta$ again denote the initial rate of growth and the effects of density-dependence, respectively, while $n$ is the degree of compensation.
When $n<1$, recruitment increases when $B$ is low, and decreases when $B$ is high, similar to the Ricker function.
When $n=1$, Eq. (\ref{eqShep}) simplifies to the Beverton-Holt (B-H) SRR, where recruitment saturates as $B$ increases.
For values of  $n>1$, recruitment behaves similarly to the Cushing function, maintaining a positive slope as $B$ increases.
The versatility of the Shepherd function comes at the cost of the additional degree of compensation parameter, which is often difficult to relate to observational data, and this has served to limit its adoption.

% Production models and Data
Using observational data, we are often unable to distinguish which model is most descriptive of the underlying dynamics.
This has been a long-standing problem: in 1982, \citeauthor{Gulland:1988vb} noted that ``in many cases, the variability of the data makes it difficult to choose between alternative mathematical models" (pg. 17).
Such variation may be a product of  environmental variability, as well as differences in life-history.
For instance, SRRs may be constrained by multiple, rather than a single compensatory event \citep{Brooks:2007cu}, and these species-specific characteristics can be controlled by many different aspects of fish reproductive biology \citep{Morgan:2011eba}. %COULD YOU SEND MORGAN Et AL TO ME PLEASE
In cases such as these, more complex models may be required, but this is at the cost of additional parameters, limiting the model's applicability to different systems. %, particularly if data are scarce.
%Moreover, the SRR, if it exists, may be unknowable.

Distinguishing between possible compensatory scenarios without assuming knowledge of the exact form the SRR would thus provide insight into the population dynamics of a fish species, without force-fitting a potentially incorrect recruitment model to observational data.   Bayesian Nonparametric techniques provide one way to estimate descriptive characteristics of stock-recruitment functions based only on the data \citep{Munch:2005dn}.

Here we present an analytical approach to determine compensatory dynamics, without assuming knowledge of the specific SRR. %though important aspects of its form are explicitly determined.
%Our generalized modeling approach \citep[sensu][]{Gross:2006wn,Gross:2009jr,Yeakel:2011p3406} permits an analytical assessment of general families of stock-recruitment models, and provides a natural interpretation for the degree of compensation parameter.
We use a generalized modeling framework \citep[\emph{sensu}][]{Gross:2006wn,Gross:2009jr,Stiefs:2010p2566,Yeakel:2011p3406,Kuehn:2010p3380} to derive relationships between the degree of compensation and the functional elasticities (the logarithmic derivative of a function, giving a measure of the change of the function relative to a change in its argument) of a continuous time generalized production model, as well as a discrete time age-structured model.
%Our analysis provides a natural interpretation for the degree of compensation parameter.
Our results demonstrate that families of SRRs can be distinguished by these functional elasticities, which can be estimated from the dynamics of perturbations in fish biomass.
We also show that some stock-recruitment families can be distinguished more easily than others, and that these differences are closely related to the stability of populations controlled by different compensatory dynamics.
%This leads to the observation that the Beverton-Holt function is qualitatively different than either the Ricker or Cushing functions; though the probability of measuring Ricker-like or Cushing-like recruitment dynamics is distinctly non-zero, we show that it is impossible to observe Beverton-Holt recruitment dynamics in nature.

\section{Methods and Analysis}
Despite the intrinsic simplifications introduced when using production or biomass dynamic models, they can offer direct insight into the mechanisms governing fish recruitment, and thus remain an important and oft-used tool in fisheries management \citep{Mangel:2002vh,Mangel:2013tu}, so we begin with them.   We then extend our results and methods to a discrete time age-structured system, and show how functional elasticities can distinguish between stock-recruitment families and provide direct insight into the stability regimes of populations with complex life histories.

\subsection{Analysis of a Generalized Stock-Recruitment Model}

In a generalized production model, we assume that biomass enlarges  according to the function $S(B)$ and shrinks according to the function $D(B)$, such that biomass changes as

\begin{equation}
\frac{\rm d}{\rm dt} B = S(B) - D(B).
\label{gen}
\end{equation}

%\noindent (B)$ and $D(B)$ are functions describing the gain and loss of biomass $B$, respectively. 
\noindent The enlargement  function $S(B)$ may be assumed to have Ricker, B-H, or Cushing recruitment dynamics, whereas $D(B)$ is often assumed to be linear, such that $D(B) = zB$, where $z$ is the rate of biomass loss due to fishing, natural mortality, or a combination thereof.
However, in many cases  we cannot assign a specific function to either $S(B)$ or $D(B)$.
%This can be problematic even when there is good mechanistic reasoning behind the use of a specific functional form, but particularly so if it is assumed that the stock-recruitment relationship cannot be determined.
Unfortunately, analysis of such a general model is not straightforward, since  the steady state solution ($B^*$, where $S(B^*) = D(B^*)$) cannot be described analytically.
In contrast, specific models present essentially the opposite problem: a steady state solution can often be computed, however the specific mathematical relationships may not accurately represent the dynamics of the population.
%Moreover, the parameters of many specific models can be difficult to interpret biologically and/or measure in the field.

The general model presented in Eq. (\ref{gen}) cannot be solved at the steady state because the functions are unknown, however we can identify the unknown steady state(s) with the variable $B^*$.
If we assume that $B^*>0$ and that the signs of the growth and loss functions are biologically meaningful, then we can normalize the system to $B^*$. 
This allows us to define a set of normalized variables and functions.  We set  $S^* = S(B^*)$ and $D^* = D(B^*)$  and define

\begin{equation}
b = \frac{B}{B^*},~~~s(b) = \frac{S(B)}{S^*},~~\mbox{and}~~ d(b) = \frac{D(B)}{D^*}.
\label{defs}
\end{equation}

This normalization procedure enables consideration of all positive steady states in the whole class of systems defined by Eq. (\ref{gen}), with the important property that at the steady state all generalized functions and variables are equal to unity ($b=1, s(1)=1, d(1)=1$)
By substituting the normalized variables into Eq. (\ref{gen}), we obtain the normalized general production model

\begin{equation}
\frac{\rm d}{\rm dt}b =\frac{S^*}{B^*}s(b) - \frac{D^*}{B^*}d(b),
\label{gen2}
\end{equation}

\noindent and under steady state conditions ($b=s(b)=d(b)=1$), this simplifies to
\begin{equation}
0 =\frac{S^*}{B^*} - \frac{D^*}{B^*}.
\label{gen3}
\end{equation}

\noindent Thus, at the steady state, the scaled growth and mortality coefficients are equivalent, allowing us to define the timescale of the system %and because Eq. (\ref{gen3}) contains only constants, these relationships hold outside of the steady state \citep{Gross:2006wn}. I DO NOT UNDERSTAND WHAT THIS MEANS THAT THE RELATIONSHIPS HOLD OUTSIDE OF THE STEADY STATE

\begin{equation}
\gamma =\frac{S^*}{B^*} = \frac{D^*}{B^*}.
\label{gamma}
\end{equation}

\noindent This parameterization is useful because $\gamma$ has a biologically relevant interpretation, and represents the biomass turnover rate. That is, for example, $S^*/B^*$ has units of production per unit time  of new biomass per unit of existing biomass.
In generalized modeling,  coefficients  such as $\gamma$ are referred to as `scale parameters' \citep{Gross:2006wn}.
Substituting $\gamma$ into Eq. (\ref{gen2}), the generalized equation is thus

\begin{equation}
\frac{\rm d}{\rm dt}b = \gamma \big(s(b) - d(b)\big).
\label{gen4}
\end{equation}

Although the normalized functions $s(b)$ and $d(b)$ are still unknown, we can assess the dynamics of Eq. (\ref{gen4}) by 
investigating the system under a small perturbation evaluated at the steady state, accomplished by taking the derivative of the normalized system.
The derivative of the right hand side of  Eq. (\ref{gen4}) is 

\begin{equation}
\left. \lambda \right|_{*} = \gamma \left( \left.\frac{\partial s(b)}{\partial b}\right|_* - \left.\frac{\partial d(b)}{\partial b}\right|_* \right) = \gamma \left(s_{\rm b} - d_{\rm b} \right),
\label{lambda}
\end{equation}

\noindent where $\lambda$ is the single eigenvalue of the system and $|_*$ indicates evaluation at the steady state $B^*$.
The system is stable if $\lambda < 0$, and unstable if $\lambda > 0$.
As $\lambda$ moves upwards toward 0, the system approaches a saddle-node bifurcation \citep{GuckHolmes,Mangel:2006wa}, a critical transition associated with the sudden appearance of a stable and unstable fixed point, changing the dynamics rapidly \citep{AleksandrovichKuznetsov:1995p2580}.
%where the stability of the system changes rapidly.

%Elasticities
The linearization of Eq. (\ref{gen4}) reveals two additional parameters, $s_{\rm b}$ and $d_{\rm b}$, which are the partial derivatives of the normalized functions $s(b)$ and $d(b)$, respectively.
%elasticities of the functions $S(B)$ and $D(B)$ with respect to the steady state $B^*$, respectively.
Partial derivatives of normalized functions are equivalent to the {\it elasticities} of the unnormalized functions \citep{Gross:2006wn,Yeakel:2011p3406}, which we show below.
In general, elasticities provide a measure of the percent change of a function relative to the percent change in its argument 
%In general, the elasticity of a function is a ratio of the percent change of the function relative to the percent change of the argument, such that

\begin{align}
{\rm Elasticity} \{F(a)\} & = \frac{a}{F(a)}\frac{\partial F(a)}{\partial a}  \nonumber\\
&= \lim_{x \rightarrow a}\frac{F(x) - F(a)}{x-a}\frac{a}{F(a)} = \lim_{x\rightarrow a} \frac{1-\frac{F(x)}{F(a)}}{1-\frac{x}{a}} \approx \frac{\% \Delta F(a)}{\% \Delta a}.
\end{align}

\noindent Elasticities are commonly used in metabolic control theory \citep{Fell:1992uh},  economics \citep{Sydsaeter:2008vr} and life history theory \citep{Horvitz:1997wy}.
%We emphasize that th but are with respect to the linearization of a normalized dynamic system, rather than a life-table matrix.

In generalized modeling, the elasticity of a function $F(X)$ with respect to its steady state $X^*$ is alternatively written as the logarithmic derivative of the function with respect to $X^*$ \citep{Yeakel:2011p3406}, and is equivalent to the partial derivative of the normalized function $f(x)$, 

\begin{equation}
f_{\rm x} = \frac{X^*}{F^*} \left.\frac{\partial F}{\partial X}\right|_{x=1} = \left.\frac{\partial \log F}{\partial \log X}\right|_{x=1} = \frac{\partial f}{\partial x}.
\end{equation}

Elasticities offer a number of advantages that are particularly useful for generalized modeling. %in a generalized modeling context.
First, an elasticity of a power-law function of the form $F(X) = aX^p$ is equal to $p$.
This can be shown by normalizing $F(X)$ to the equilibrium $X^*$, and taking the derivative at the steady state, such that

\begin{equation}
f_{\rm x} = \left.\frac{\partial f}{\partial x} \right|_* = \left. \frac{\partial}{\partial x} \frac{aX^p}{{aX^*}^p}\right|_*=\left.\frac{\partial}{\partial x} x^p\right|_* = p. \nonumber
\end{equation}

\noindent For instance, if the function $D(B)= zB$ and $z$ is a constant, then the elasticity is equal to unity; if the function is quadratic, the elasticity is equal to 2; for constant functions, the elasticity is equal to 0.
For more complex functions, the value of the elasticity may change with the value of the steady state (see below).
Importantly, the elasticities of functions governing the time-evolution of an animal population are representative of the environmental conditions present during measurement.
Thus elasticities are not defined with respect to  unmeasurable biological conditions that serve to bound traditional functional relationships, such as half-maximum values or growth rates at saturation  \citep{FELL:1985p3430,Fell:1992uh}.
%Compared to the $\alpha$ and $\beta$ variables defined in traditional stock-recruitment relationships (Eqs \ref{eqRick},\ref{eqBH},\ref{eqCush}), elasticities are not defined with respect to unrealistic biological conditions, such as half-maximum values or growth rates at saturation \citep{FELL:1985p3430,Fell:1992uh}.

\subsection{Relating functional elasticities to the degree of compensation}

The degree of compensation in the Shepherd function (Eq. \ref{eqShep}) is controlled by the parameter $n$: if $n<1$ the function is Ricker-like, if $n>1$ the function is Cushing-like, and if $n=1$ it is equivalent to the B-H function \citep{Shepherd:1982p3428}.
In a generalized modeling framework, the degree of compensation is related directly to the functional elasticity.
Given that $B^*$ is large enough to experience density-dependent effects, if the elasticity of growth $s_{\rm b} < 0$ the population grows according to a Ricker-like function, if $s_{\rm b} > 0$ the population grows according to a Cushing-like function, and if $s_{\rm b} \rightarrow 0$ the population grows according to the B-H function.
Thus, $n$ and $s_{\rm b}$ are closely related, which can be shown by mapping the Shepherd function (Eq. \ref{eqShep}) to the generalized model (Eq. \ref{gen4}), where

\begin{align}
s(b) = \frac{S(B)}{S(B^*)} &= \frac{\alpha B}{1 + \beta B^{1/n}} \cdot \frac{1 + \beta B^{*1/n}}{\alpha B^*} ,\nonumber \\
& = \frac{1 + \beta B^{*1/n}}{1 + \beta B^{*1/n}b^{1/n}}b. \nonumber
\end{align}

\noindent  The elasticity of growth is

\begin{equation}
s_{\rm b} = 1 - \frac{\beta B^{*1/n}}{n(1+ \beta B^{*1/n})}.
\label{map1}
\end{equation}

\begin{figure}[h!]
   \centering
   \includegraphics[width=0.75\textwidth]{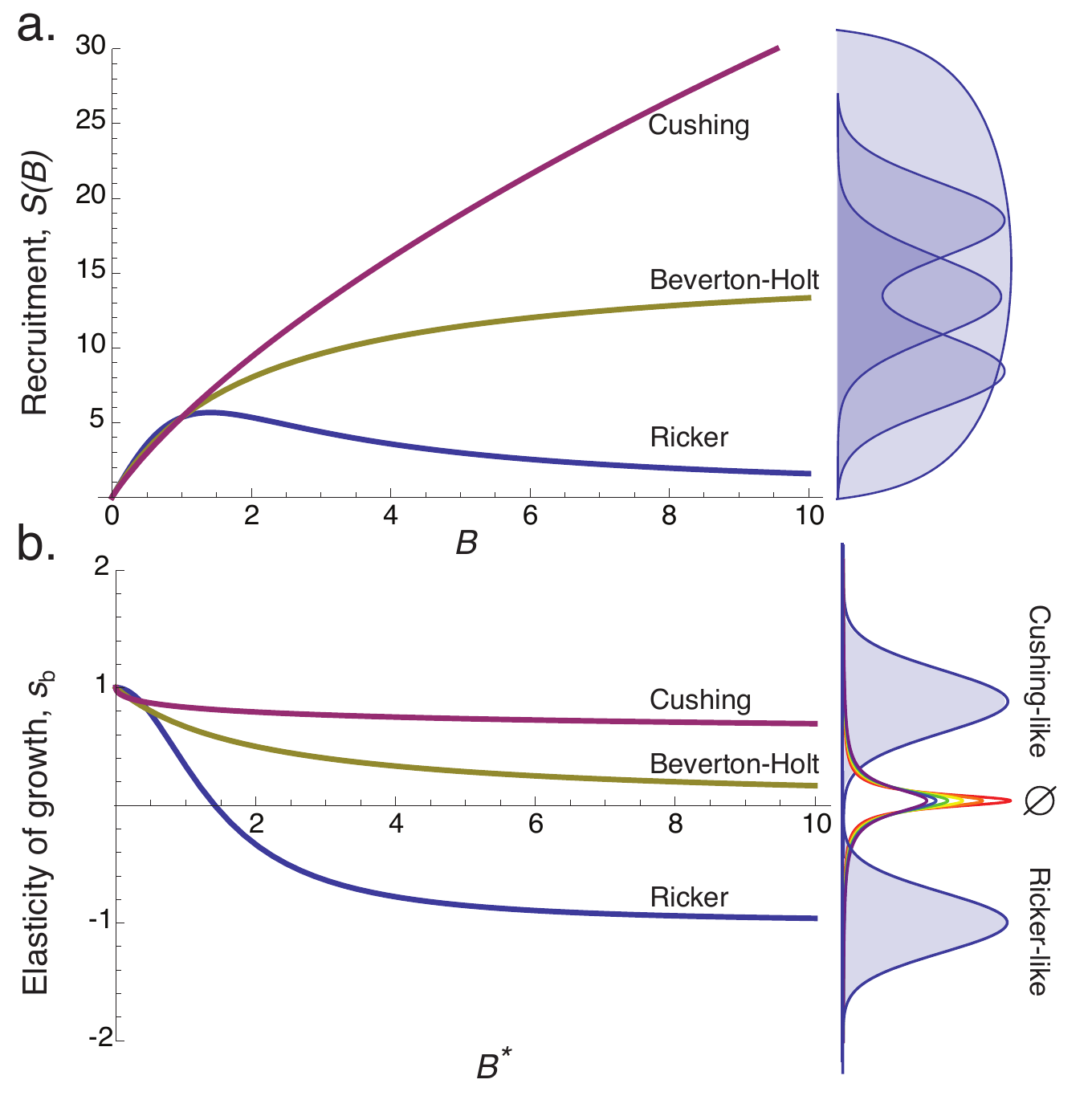}
      \caption{
      a. Recruitment $S(B)$ as a function of spawning stock biomass $B$.
      b. The elasticity of growth $s_{\rm b}$ as a function of the steady state spawning stock biomass $B^*$.
      The distributions to the right represent potential measurements of recruitment (a.) and the elasticity of growth (b.) for Ricker, B-H, and Cushing recruitment functions.
      Although measurements for SRRs overlap in (a.), the elasticities of the Ricker- and Cushing-like functional families can be represented by non-overlapping intervals. Ricker: $s_{\rm b} \in [-\infty,0)$; Cushing: $s_{\rm b} \in (0,\infty]$), whereas the B-H function is the boundary between Ricker- and Cushing- functional families, such that $s_{\rm b}  \in \varnothing$.
       }
      \label{figcompdist}
\end{figure}

\noindent Eq. (\ref{map1}) shows that the elasticity of growth  depends on both the steady state biomass, as well as the degree of compensation, enabling direct comparisons between Ricker-like, Cushing-like, and B-H functions and their corresponding elasticities.
%By comparing the Shepherd function (Eq. \ref{eqShep}) and its corresponding elasticity, the relationships posed above are shown (Fig. \ref{figcomp}).
For example, as $B^*$ increases, if $n > 1$ (Cushing), then $s_{\rm b} > 0$; if $n < 1$ (Ricker), then $s_{\rm b} < 0$; if $n = 1$ (B-H), $s_{\rm b} \rightarrow 0$ (Fig. \ref{figcompdist}). 
(This is more readily apparent if the quantity $(1/\beta B^*{}^{1/n})(1/\beta B^*{}^{1/n})^{-1}$ is factored into the rightmost term of Eq. \ref{map1}.)
Because the value of the elasticity holds for any function $S(B)$, this generalization is not isolated to the Shepherd equation, but extends to any function with degrees of compensation that can be categorized as `Ricker-like', `Cushing-like', or saturating. %that has Ricker-like, Cushing-like, or B-H behavior.
Thus, when density-dependent effects are present, if the value of the elasticity $s_{\rm b}$ can be determined, a general functional family can be assigned to the observed recruitment dynamics.
This is a key relationship, because assignment of the functional family does not depend on the specific architecture of a given function.

If we assume that recruitment follows a  Shepherd function, the degree of compensation can be determined directly if the elasticity $s_{\rm b}$ is
%For instance, if mortality is governed by a linear function ($D(B) = zB$, such that the elasticity of mortality $d_{\rm b} = 1$), then

\begin{equation}
s_{\rm b} = \frac{\gamma + \alpha(n-1)}{\alpha n},~~\mbox{or alternatively,}~~n = \frac{\gamma - \alpha}{\alpha(s_{\rm b} - 1)},
\end{equation}

\noindent where as before, $\gamma$ is the biomass turnover rate, and $\alpha$ is the recruitment rate at low biomass. %with units of recruits$\cdot$biomass${}^{-1}$.
%JUSTIN -- SHOULD WE EXPLICITLY MENTION Z AND SHOW WHERE IT IS?
From this relationship, we see that if $s_{\rm b}<1$, $\gamma$ is constrained to vary between 0 and $\alpha$ if $n$ is to remain positive.
Because $\gamma$ is the biomass loss rate, it is evident that values greater than $\alpha$ (the maximum growth rate independent of density dependent effects) imply extinction of the population.

%Data & 
The discrimination of different governing functional forms (or families of functional forms) from observational data typically requires measures of statistical best fit using multiple years of stock-recruitment data (Munch et al 2005).
Because these data are often highly variable and complicated by changes in birth and death rates over long timespans, distinguishing between functional forms can be problematic (Fig. \ref{figcompdist}a).
However, because the elasticities of alternative functional families have non-overlapping ranges, they may be useful for determining the effects of density dependence on recruitment (Fig. \ref{figcompdist}b).
Moreover, because the sign of the elasticity can differentiate between competing functional families, the determination of functional family from the elasticity of growth may be relatively error-tolerant.

%Measure Zero
The relationship between the elasticity of growth and the degree of compensation suggests that the Cushing SRR and Ricker SRR are qualitatively different than the B-H SRR.
The reasoning for this is straightforward: the elasticity of growth is a continuous variable, and recruitment following the B-H function is defined by the elasticity $s_{\rm b} = 1$, whereas Cushing-like and Ricker-like functions have elasticities that span a range of values.
Mathematically, the elasticities of the Ricker- and Cushing-like functional families can be represented by non-overlapping intervals (Ricker: $s_{\rm b} \in [-\infty,0)$; Cushing: $s_{\rm b} \in (0,\infty]$), whereas because the B-H function represents the boundary between the Ricker- and Cushing- functional families, it is a measure zero, or null set (B-H: $s_{\rm b}  \in \varnothing$).
%Because the probability that a continuous variable is equal to a single value is always zero, it is evident that recruitment following the B-H functional form cannot exist in nature (Fig. \ref{figcompdist}b).

%%%
\subsection{Measuring elasticities from time-series}

We have shown that the degree of compensation can be calculated if the elasticity of growth is known.
There exists a large body of literature in metabolic control theory for measuring elasticities in nature \citep[which typically consist of experimental manipulations;][]{Fell:1992uh}, however these tools are not always appropriate for obtaining measurements from animal populations in the wild.
We  now show  that the elasticity of growth can be measured from relatively small perturbations in fish biomass, and we provide a basic example using simulated data.

To begin, we consider single-species dynamics, where ${\rm d}X/{\rm dt} = F(X)$.  We  define deviation from the steady state, such that the population size at time $t$ is some distance away from the equilibrium $X^*$ as $ \xi(t) = X(t) - X^*$.
Then to first order

\begin{equation}
\frac{\rm d}{\rm dt}\xi(t) \approx F'(X^*)  \xi.
\label{dev}
\end{equation}

% where $F'(X^*)$ is the time-derivative of $F(X^*)$. 
\noindent For a single-species system, we observe that $F'(X^*)$ is also the single eigenvalue of the system, $\lambda_f$, and we use the subscript $f$ to distinguish the eigenvalue in this example from the eigenvalue $\lambda$ defined for the production model. %THIS IS NOT THE SAME EIGENVALUE AS BEFORE, HOWEVER.  SHOULD WE USE A DIFFERENT SYMBOL?
Integrating Eq. (\ref{dev}), we find that $\xi(t) =  \xi_0{\rm e}^{\lambda_f t}$ where $\xi_0$ is the initial deviation and $X(t) = X^* + \xi_0{\rm e}^{\lambda_f t}$.
Thus, the eigenvalue of a single-species system is equivalent to the rate of relaxation to the steady state of the population trajectory after a small perturbation if $\lambda_f < 0$.

Our generalized analysis of Eq. (\ref{gen}) shows that $\lambda = \gamma(s_{\rm b} - d_{\rm b})$.
For now, we will assume that $\lambda$ can be measured.
To determine which of the three functional families depicted in Fig. (\ref{figcompdist}a) drive recruitment dynamics, we must determine the elasticity of growth, where $s_{\rm b} = \lambda/\gamma + d_{\rm b}$.
%Thus, we can distinguish between qualitative stock recruitment models by relating the rate of relaxation to the steady state to both the biomass turnover rate and the elasticity of mortality.
If mortality is assumed to be governed by a linear function, such that $d_{\rm b} = 1$, then the criteria are simply defined by comparing the magnitude of the relaxation rate, $\lambda$, to the timescale of the system, $\gamma$ (Table \ref{tab1}).
If we assume that the steady state is stable ($\lambda < 0$), recruitment is driven by a Ricker-like function if $\lambda < -\gamma$, recruitment is driven by the B-H function if $\lambda = -\gamma$, and recruitment is driven by a Cushing-like function if $\lambda > -\gamma$ (Fig. \ref{figcompdist}a).
Because we do not presume to know the exact architecture of the stock-recruitment function, these relationships are predictive of general families of models. %and these designations should be robust to measurement error.
%In other words, because of the generality of the questions asked, these criteria are robust to measurement/process noise.
If we assume that growth is governed by the Shepherd function, the general relationship between the degree of compensation and the relaxation rate is  (cf. Eqn 15)

\begin{equation}
n = \gamma \frac{\gamma - \alpha}{\alpha (\lambda+\gamma d_{\rm b}-\gamma)}.
\label{gen_n}
\end{equation}

\noindent which can be simplified further assuming that mortality is governed by a linear function, such that % ($D(B) = zB$, where the elasticity of mortality $d_{\rm b} = 1$)

\begin{equation}
n = \gamma \frac{\gamma - \alpha}{\alpha \lambda}.
\label{gen_n}
\end{equation}

%\noindent where different values of $\gamma$ result in relatively modest changes in $n$, particularly for systems far from the bifurcation $\lambda = 1$.

\begin{figure}[h!]
   \centering
   \includegraphics[width=0.4\textwidth]{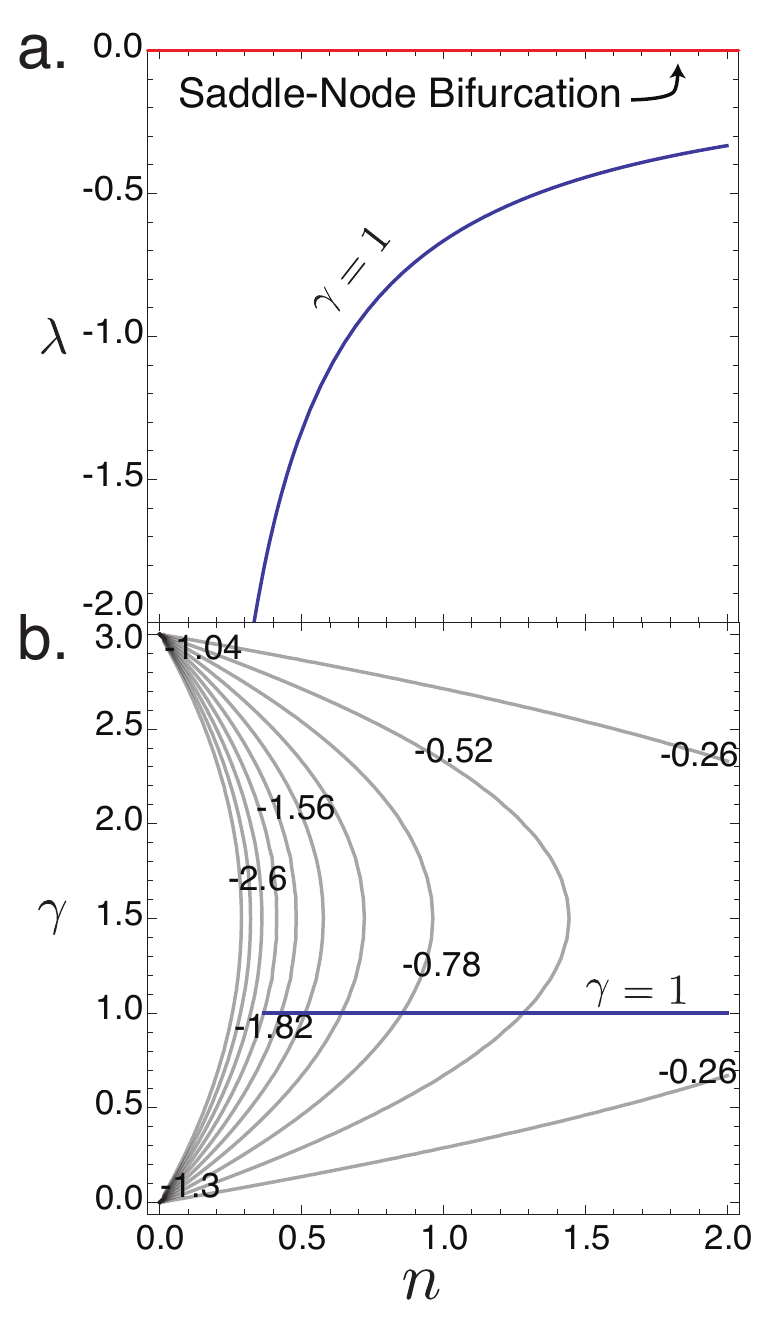}
      \caption{
       a. The rate of relaxation to the steady state $\lambda$ vs. the degree of compensation $n$ for a biomass turnover rate $\gamma =1$.
       The red line at $\lambda = 0$ denotes a saddle-node bifurcation below which the system is stable, and above which the system is unstable.
       b. The biomass turnover rate $\gamma$ as a function of the degree of compensation $n$ and the rate of relaxation $\lambda$ (contour lines). 
       The trajectory shown in (A.) is denoted by the blue line.
       Values of $0<\gamma < (\alpha=3)$ result in stable dynamics, and only Cushing-like functions, where $n>1$ can result in values of $\lambda$ close to the saddle-node bifurcation.
       }
      \label{figShep}
\end{figure}

As the system approaches the saddle-node bifurcation at $\lambda = 0$, small errors in $\lambda$ are likely to generate large errors in the degree of compensation (Eq. \ref{gen_n}, Fig. \ref{figShep}), such that Ricker-like SRRs  result in measurements that are more error-tolerant than Cushing-like SRRs.
Moreover, because an elasticity of growth $s_{\rm b}<1$ produces dynamics with a single non-trivial stable steady state (assuming the elasticity of mortality $d_{\rm b}=1$), only Cushing-like SRRs can come close to the saddle-node bifurcation at $\lambda = 0$. %, resulting in unstable dynamics.
The rate at which the saddle-node bifurcation is reached as $n$ increases is contingent on the biomass turnover rate $\gamma$, where turnover rates intermediate to 0 and $\alpha$ approach the bifurcation more slowly.
If the turnover rate is greater than $\alpha$, $\lambda>0$ and the system becomes unstable. %, thus bounding the values of $\gamma$ and $\alpha$ in stable systems.
%If the turnover rate is less than $\alpha$, and mortality is linear, the system is always stable, as the Shepherd function cannot have an elasticity of growth $>1$.

\subsection{Estimating the degree of compensation from fluctuations in fish biomass}

\begin{figure*}[h]
   \centering
   \includegraphics[width=0.75\textwidth]{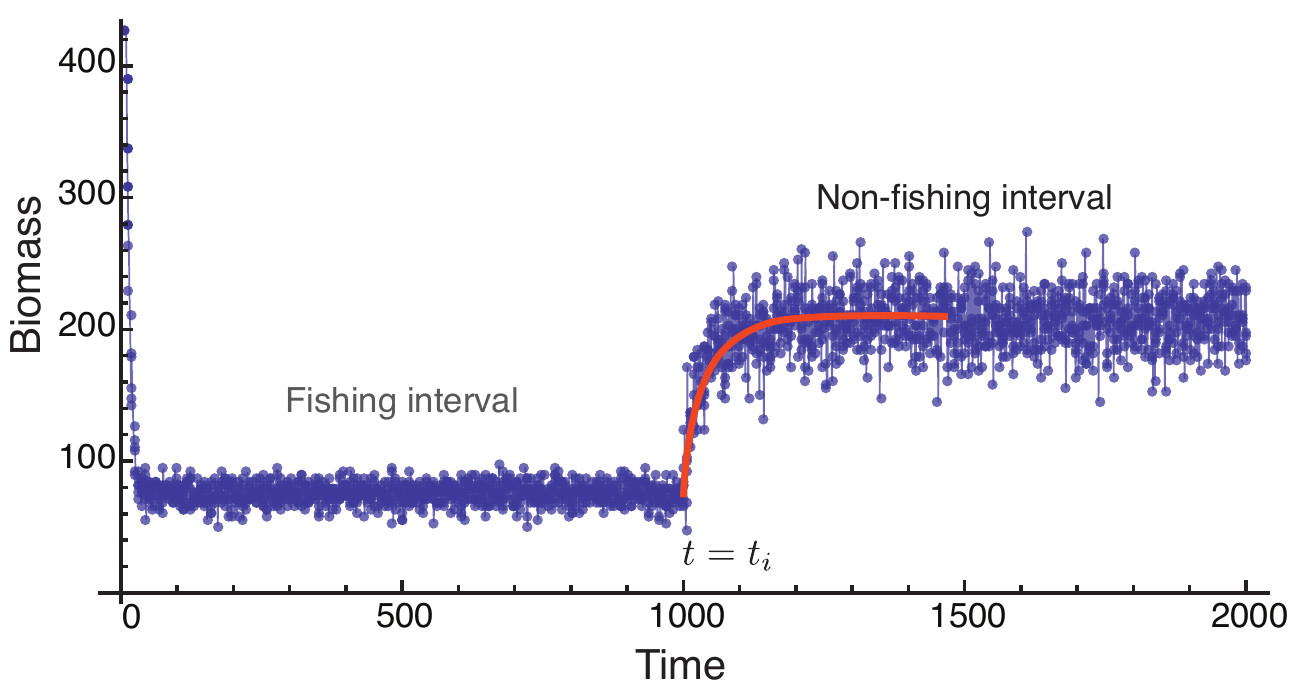}
      \caption{
      An example of the transition from a fishing to a non-fishing interval used to measure the rate of relaxation $\lambda$ from time-series data.
      The non-fishing interval is initiated at $t=t_i$, and biomass values immediately after $t_i$ can be used to find the maximum likelihood estimate for $\lambda$.
      The best-fit trajectory using the likelihood technique is shown in orange.
      }
      \label{figTS}
\end{figure*}

We have derived a relationship between the degree of compensation $n$ and the elasticity of growth $s_{\rm b}$, and have shown how - in principle - elasticities could be measured from short-term fluctuations in time-series data.
To elaborate this idea, we constructed a stochastic model with growth following the Shepherd function and mortality due to both natural causes $M$ and fishing $F$, coupled with observation error. %JUSTIN: FISHERY SCIENTISTS USE UPPER CASE FOR THESE AND WE SHOULD TOO
We perturbed the system at time $t = t_i$ by eliminating the fishing mortality term (the end of a fishing period) until a steady state was reached at the terminal time $t = T$.
We included normally distributed observation error $\tilde{P}$  with mean zero and standard deviation $\sigma$. 
Accordingly, observations of fish biomass $B_{\rm obs}(t)$ change are then

\begin{align}
& \frac{\rm d}{\rm dt}B = \frac{\alpha B}{1 + \beta B^{1/n}} - (M + \delta F)B, \nonumber \\ 
& B_{\rm obs}(t) = B(t) + \sigma \tilde{P}, \nonumber
\end{align}

\noindent where $\delta$ controls fishing mortality. %and $B_{\rm obs}$(t) is the observed biomass at time $t$.
During the fishing time interval $t_0 \leq t < t_i$, $\delta = 1$; during the non-fishing time interval $t_i \leq t < T$, $\delta = 0$ (e.g. Fig. \ref{figTS}).

Given the Gaussian assumption about the observation error, we assume that the system trajectory behaves as %$B_{\rm obs}(t) = c (1 - {\rm e}^{-\lambda t}) + \sigma {\rm Gauss}\{0,1\}$, or more succinctly 
$B_{\rm obs}(t) \sim {\rm N}\{c (1 - {\rm e}^{-\lambda t}),\sigma\}$ close to the steady state. %where $c$ is the biomass at the initiation of the perturbation.
The stochastic trajectory  thus depends on the unknown variables $c$, $\lambda$, and $\sigma$, which we determine using a likelihood approach where $k$ is the number of observations from the end of the fishing period until the trajectory reaches its steady state in the absence of fishing at $t=T$. %and $\mu = c (1 - {\rm e}^{-\lambda t})$.
%, and the derivative of the log-likelihood can be determined to calculate the maximum likelihood estimates of $c,~\lambda,$ and $\sigma$.
This problem can be simplified, as the variables $c$ and $\sigma$ can be written in terms of $\lambda$ to obtain the log-likelihood \citep{Hilborn:1997ds}

%\begin{equation}
%{\mathscr L}(c,\lambda,\sigma) = \prod_{t=t_i}^{T}\frac{1}{2\pi\sigma^2} \left[-\frac{(B(t) - \mu(t))^2}{2\sigma^2}\right], \nonumber
%\label{like}
%\end{equation}

\begin{equation}
{\rm log}{\mathscr L}(\lambda) = -\frac{k}{2}{\rm log}(2\pi)-\frac{k}{2}{\rm log}\left\{\frac{1}{k}\left(\sum_{t=t_i}^TB(t) -\frac{\sum_{t=t_i}^TB(t)(1-{\rm e}^{-\lambda t})}{\sum_{t=t_i}^T(1-{\rm e}^{-\lambda t})}\right)^2\right\} - \frac{k}{2},
\label{loglike}
\end{equation}

\noindent which we use to find the maximum likelihood estimate for the eigenvalue $\lambda_{\rm MLE}$.

%Determination of functional forms
We aim to discriminate between different families of functional forms using the maximum likelihood estimate for the rate at which a population trajectory returns to its steady state after a perturbation.
If the rate of relaxation is known, the degree of compensation can be calculated from Eq. (\ref{gen_n}).
To determine the accuracy of our model estimates across different degrees of observation noise, we calculated $\lambda_{\rm MLE}$ as a function of the coefficient of variation (${\rm CV}=\sigma/B^*$) for three compensation scenarios: Ricker ($n=0.5$), Cushing ($n=1.5$), and B-H ($n=1$).
After estimating $\lambda_{\rm MLE}$, we calculated the degree of compensation, $n_{\rm MLE}$ from Eq. (\ref{gen_n}), and determined the probability that $n$ was correctly distinguished with respect to alternative stock-recruitment functions (Fig. \ref{figProb}a,b), as well as the probability that the functional family was correctly identified (Fig. \ref{figProb}a,c).

\section{Results}

The analytical relationship between the degree of compensation $n$ and the elasticity of growth $s_{\rm b}$ (Eq. \ref{gen_n}) suggests that populations growing in accordance to Ricker-like functions should be less difficult to measure accurately than those growing in accordance to Cushing-like functions (Fig. \ref{figShep}a).
In general, our simulation experiment showed that the rate of relaxation can be estimated from moderately noisy data, and that  there were large differences in the measurement accuracy for different functional families.
The estimated rate of relaxation $\lambda_{\rm MLE}$, and by transformation $n_{\rm MLE}$, is estimated more accurately for Ricker-like and B-H functions than for Cushing-like functions (Fig. \ref{figProb}a).
We note that the mean value of our estimates always diverged from the set value of $n$ because the rate of return equation is only accurate at values very close to $B^*$ and is therefore a necessarily crude estimate of the solution to $B_{\rm obs}(t)$.

Despite differences in measurement accuracy, the probability that the correct stock-recruitment function was distinguishable from the other SRRs declined approximately linearly for both Ricker and Cushing models after ${\rm CV} = 0.15$, while that for the B-H model declined nonlinearly (Fig. \ref{figProb}b).
The B-H model was more difficult to distinguish because estimates of $n$ overlapped values for both neighboring models.
However, this comparison is somewhat arbitrary, and the more important question relates to the probability that the functional family is correctly determined with respect to other potential families of functions.
Our results showed that the probably of correctly determining the functional family from $\lambda_{\rm MLE}$ remained relatively high as CV increased (Fig. \ref{figProb}c) for both Ricker-like and Cushing-like functions.
The probably that Ricker-like functions were correctly distinguished was generally greater than $0.6$ for ${\rm CV} \leq 0.30$.
The same probability was greater than $0.8$ for $\rm{CV} \leq 0.5$ for Cushing-like functions, due primarily to the greater range of $n$ for the Cushing functional form.
Because the B-H function is boundary between functional families (see Methods), the probability that it was correctly distinguished was always 0.

% Extending to age-structured models.
\section{An Example With Age Structure}
Production models effectively summarize the recruitment dynamics of fish populations, and in some cases can provide robust measures of fisheries reference points \citep{MacCall:2002id,Mangel:2010tp,Mangel:2013tu}.
However, the influence of age-related differences in growth and mortality can have large effects on the dynamics of fish populations \citep{Mangel:2006tm,Shelton:2011eq}.
In this section we build upon our prior results and expand the generalized modeling schema to discrete time, age-structured models.
%Moreover, we present our age-structured extension using a discrete-time derivation for generalized modeling approaches, which has not been published previously.
The extension of generalized modeling to discrete time systems is useful in its own right, as it provides a method for the dynamical analysis of whole classes of discrete time models \citep[\emph{sensu}][]{Gross:2006wn}. %, instead of those with specific functional forms.
First, we briefly illustrate an extension of the generalized modeling approach to an age-structured discrete time system.
Second, we show how the degree of compensation in an age-structured system is related to the elasticities of growth and finish by showing how measurements of elasticities in the age-structured model provide important insight into system stability.

We consider an age-structured model where the number of recruits $X(t+1)$ is governed by spawning stock biomass, $B_s(t)$, depending on the degree of compensation $n$.
The number of individuals in the mature age class $Y$ is the sum of returning adults and incoming recruits, where adult mortality is given by $M_y$ and recruit mortality is given by $M_x$. 
It follows that spawning stock biomass is calculated by the number of mature individuals times the average mass of individuals $W_y$.
The age-structured model is thus 

\begin{align}
\label{age1}
X(t+1) &= F(B_s,t) ~~~~~~~~~~~~~~~~ = \frac{\alpha B_s(t)}{1+ \beta B_s(t)^{1/n}}, \nonumber\\
Y(t+1) &= G(X,t) + H(Y,t) ~= X(t){\rm e}^{-M_x} + Y(t) {\rm e}^{-M_y}, \nonumber \\
B_s(t+1) &= K(Y,t) ~~~~~~~~~~~~~~~~~=Y(t)W_y.
\end{align}

%ALTERNATIVELY WE COULD WRITE
%
%\begin{align}
%\label{age1}
%X(t+1) &= F(B,t) - D(B,t) = \frac{\alpha B(t)}{1 + \beta B(t)^{1/n}} +X(t)e^{-f_X}{\rm e}^{-M_x} , \nonumber\\
%Y(t+1) &= G(X,t)  ~~~~~~~~~~~~~= X(t)(1-e^{-f_X}){\rm e}^{-M_x} + Y(t)e^{-M_y}, \nonumber\\
%B(t+1) &= H(Y,t)  ~~~~~~~~~~~~~= Y(t)W_y,
%\end{align}

\noindent We can determine the steady state condition $X^*$ (where $X(t+1) = X(t)$) in terms of spawning stock biomass $B_s^*$ because at the steady state $B_s^* = Y^*W_y$, such that

\begin{equation}
B_s^* = X^*\frac{{\rm e}^{-M_x}W_y}{1-{\rm e}^{-M_y}} = X^*W_f^*.
\end{equation}

\noindent The primary difference between the age-structured and production models is that mortality is not assumed to occur simultaneously with recruitment, and this yields dynamics that diverge strongly from those predicted by the production model.
We note that this 3-dimensional model can be slightly modified and collapsed such that $B_s(t) = Y(t)W_y$, and this has little effect on the qualitative dynamics.

The normalization of the age-structured system is analogous to the normalization of the production model (Eq. \ref{gen2}), however because the age-structured system is composed of difference rather than differential equations, the steady state condition requires that the scale parameters are defined differently than before.
For example, $y(t+1) = (G^*/Y^*)g(x,t)+(H^*/Y^*)h(y,t)$ is defined at the steady state $1 = G^*/Y^* + H^*/Y^*$, such that we define the ratio of incoming recruits to the abundance of the mature age-class $\gamma_y = G^*/Y^*$ and the ratio of returning adults to the abundance of the mature age-class $(1-\gamma_y) = H^*/Y^*$.
These coefficients are thus the proportional contributions of recruit and mature age-classes to spawning stock biomass at the steady state.
The generalized system is then 

\begin{align}
x(t+1) &= \gamma_x f(b,t), \nonumber\\
y(t+1) &= \gamma_yg(x,t) + (1-\gamma_y)h(y,t), \nonumber\\
b(t+1) &= \gamma_bk(y,t).
\end{align}

\noindent We can immediately simplify the problem by observing that the scale parameters for recruits and biomass can be reduced to

\begin{align}
\gamma_x &= \frac{F^*}{X^*} = \frac{\alpha W^*_f}{1+\beta(W^*_fX^*)^{1/n}} = 1, \nonumber\\
%\gamma_y &= \frac{G^*}{Y^*} = 1-{\rm e}^{-M_y} \nonumber \\
%(1-\gamma_y) &= \frac{H^*}{Y^*} = {\rm e}^{-M_y} \nonumber \\
\gamma_b &= \frac{K^*}{B_s^*} = 1.
\end{align}

%\noindent Because the functions $F$ and $K$ are solely responsible for the dynamics of $X$ and $B_s$ from time $t$ to $t+1$, at the steady state $\gamma_x=\gamma_b=1$.
%By comparison, the steady state condition of the adult age-class must balance the ratios of incoming adults to the returning adults, and our specific model demands that $\gamma_y=1-{\rm e}^{-M_y}$.

For the production model, elasticities were defined with respect to the linearized system (Eq. \ref{lambda}).
Because the age-structured system is multi-dimensional, the linearization is defined by the Jacobian matrix evaluated at the steady state $\mathbf{J}|_*$, where each element is defined by the partial derivative of each differential equation with respect to each variable.
The elasticities of the generalized system can then be calculated, such that

%\begin{align}
%\frac{\partial f(b,t)}{\partial x} &= 0~~ \frac{\partial f(b,t)}{\partial y} = 0~~\frac{\partial f(b,t)}{\partial b} = \frac{\alpha W_f^*(n-1) + 1}{\alpha W_f^* n} \nonumber\\
%\frac{\partial g(x,t)}{\partial x} &= 1~~\frac{\partial g(x,t)}{\partial y} = 0~~\frac{\partial g(x,t)}{\partial b} = 0 \nonumber\\
%\frac{\partial h(y,t)}{\partial x} &= 0~~\frac{\partial h(y,t)}{\partial y} = 1~~\frac{\partial h(y,t)}{\partial b} = 0 \nonumber\\
%\frac{\partial k(y,t)}{\partial x} &= 0~~\frac{\partial k(y,t)}{\partial y} = 1~~\frac{\partial k(y,t)}{\partial b} = 0. \nonumber\\
%\end{align}

%\noindent The Jacobian is then

\begin{equation} 
\mathbf{J}|_* =
%\left(
%\begin{array}{ccc}
%\gamma_x \left.\frac{\partial f}{\partial x}\right|_* & \gamma_x\left.\frac{\partial f}{\partial y}\right|_* & \gamma_x\left.\frac{\partial f}{\partial b}\right|_* \\
%\gamma_y \left.\frac{\partial g}{\partial x}\right|_* & \gamma_y\left.\frac{\partial g}{\partial y}\right|_* & \gamma_y\left.\frac{\partial g}{\partial b}\right|_* \\
%\gamma_b \left.\frac{\partial h}{\partial x}\right|_* & \gamma_b\left.\frac{\partial h}{\partial y}\right|_* & \gamma_b\left.\frac{\partial h}{\partial b}\right|_* \\
%\end{array}
%\right) =
\left(
\begin{array}{ccc}
\left.\frac{\partial F}{\partial X}\right|_* & \left.\frac{\partial F}{\partial Y}\right|_* & \left.\frac{\partial F}{\partial B_s}\right|_* \\ 
\left.\frac{\partial G+H}{\partial X}\right|_* & \left.\frac{\partial G+H}{\partial Y}\right|_* & \left.\frac{\partial G+H}{\partial B_s}\right|_* \\ 
\left.\frac{\partial K}{\partial X}\right|_* & \left.\frac{\partial K}{\partial Y}\right|_* & \left.\frac{\partial K}{\partial B_s}\right|_* \\
\end{array}
\right) =
\left(
\begin{array}{ccc}
0 & 0 & f_{\rm b} \\
\gamma_y & (1-\gamma_y) & 0 \\
0 & 1 &0
\end{array}
\right),
\label{Jac}
\end{equation}

\noindent and for the specific age-structured model, $\gamma_y=1-{\rm e}^{-M_y}$, and 

\begin{equation}
f_{\rm b} =\frac{\alpha W_f^*(n-1)+1}{\alpha W_f^* n}.
\label{growth_age}
\end{equation}

% Stability
\noindent The Jacobian matrix  determines the stability of the system; we solve for the eigenvalues that satisfy the characteristic equation ${\rm Det}(\mathbf{J}|_*-\lambda \mathbf{I})=0$, where $\mathbf{I}$ is the identity matrix.
From Eq. (\ref{Jac}), the characteristic equation is $f_{\rm b} \gamma_y+\lambda^2-\gamma_y\lambda^2-\lambda^3=0$, which yields three distinct eigenvalues, and though the solutions for these eigenvalues are large and unwieldy, they can be easily derived with algebraic computing languages such as Maple or Mathematica.

\begin{figure*}[h]
   \centering
   \includegraphics[width=1\textwidth]{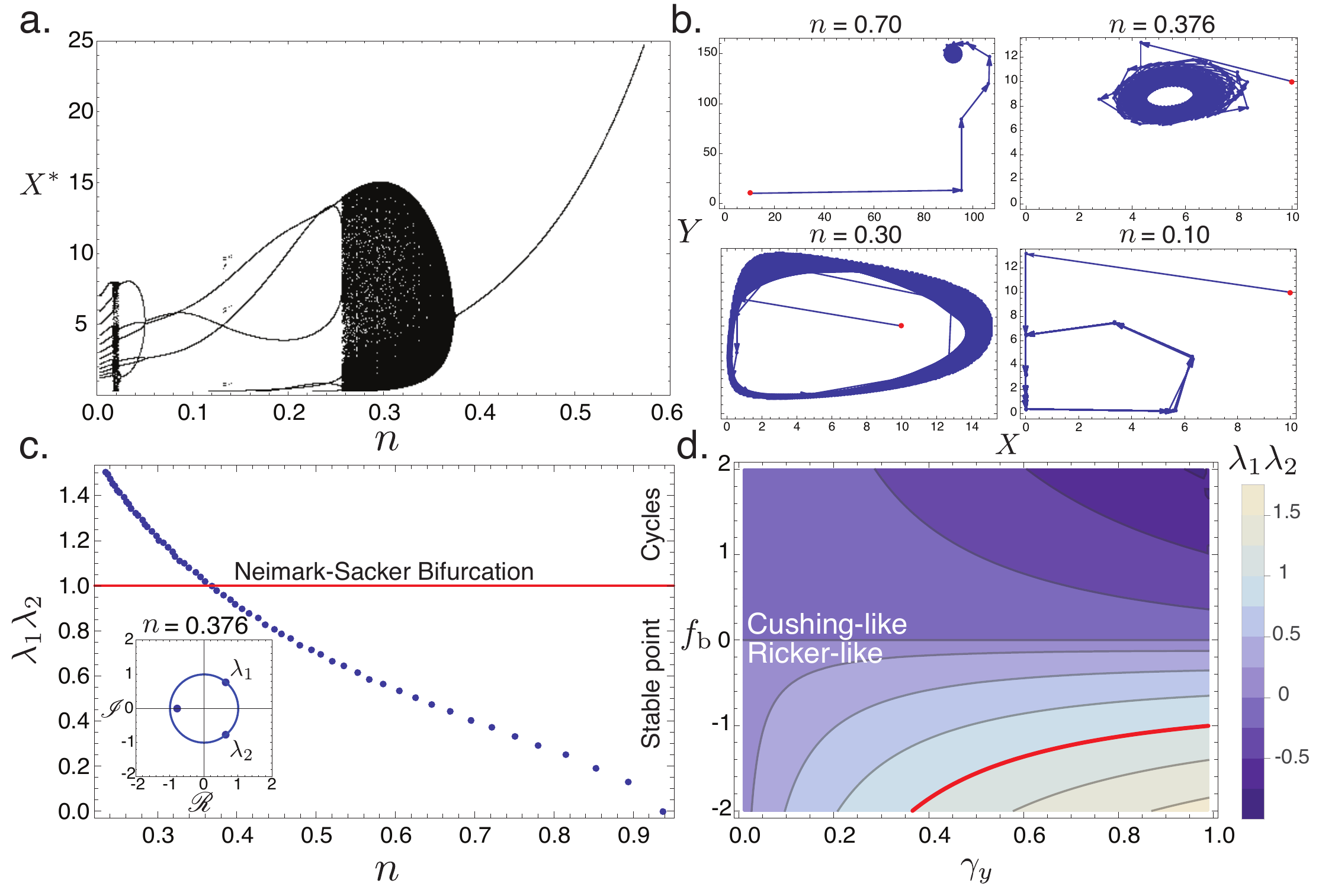}
      \caption{
      a. Bifurcation diagram showing the onset of cycles followed by multi-period oscillations for the age-structured model as the degree of compensation lowers beyond $n=0.376$.
      b. Examples of the corresponding dynamics where $n=0.70$, $n=0.376$, $n=0.30$, and $n=0.10$.
      c. Values of the test-function $\lambda_1\lambda_2$ across different values of $n$ for the specific model. A Neimark-Sacker bifurcation exists at $\lambda_1\lambda_2=1$, which is crossed at $n=0.376$. This condition exists when two complex conjugate eigenvalues cross the unit circle on the complex plane (inset).
     d. Numerically estimated values for the test-function $\lambda_1\lambda_2$, given the elasticity of growth $f_{\rm b}$ and the ratio of incoming recruits to the mature age-class $\gamma_y$. 
     The red contour denotes the Neimark-Sacker bifurcation condition; systems below this contour have cyclic dynamics.
      }
      \label{figest_age}
\end{figure*}

Simulation of the age-structured system (where $\alpha = 8,~\beta = 1/80,~M_x=0.2,~M_y=0.7$, and $W_y=2$) across a range of values for the degree of compensation reveals a single steady state for $n>0.376$.
For $n<0.376$, stable cycles emerge, which in turn give rise to five-period cycles for lower values of $n$ (Fig. \ref{figest_age}a,b).
In discrete time systems, the emergence of cyclic conditions can result from crossing a Neimark-Sacker bifurcation \citep[cf.][]{Guill:2011bf,Guill:2011vx}, which occurs when a pair of complex conjugate eigenvalues cross the unit circle on the complex plane.
If $\lambda_1$ and $\lambda_2$ are the complex conjugate eigenvalue pair, the test-function for this condition is $\lambda_1\lambda_2 = 1$ \citep{AleksandrovichKuznetsov:1995p2580}.
Using solutions for $\lambda_1$ and $\lambda_2$ from the characteristic equation, we numerically determined that a supercritical Neimark-Sacker bifurcation is crossed at $n = 0.376$ (Fig. \ref{figest_age}a,c).
Supercritical Neimark-Sacker bifurcations yield stable closed invariant curves, such that local trajectories initiated interior and exterior to the cycle are attracted to the curve \citep[cf. Fig. \ref{figest_age}b;][]{AleksandrovichKuznetsov:1995p2580}.
%The age-structured model thus effectively inverses the relationship between stability regimes and the degree of compensation, and 
Predictions of population dynamics are thus possible, but only if the degree of compensation, in addition to the other parameters, is known.
As before, the specific age-structured model introduces strong assumptions regarding functional forms, and these assumptions may not hold (or be conducive to measurement) in many situations.

Because the degree of compensation is related directly to the elasticity of growth (Eq. \ref{growth_age}), we can use the generalized age-structured system to gather direct insight into the potential dynamics applicable to any class of models substituted into the general functions $F(B)$, $G(X)$, $H(Y)$, and $K(Y)$.
Although the test-function for the Neimark-Sacker bifurcation is not analytically tractable (even for the generalized system), we can numerically simulate the relationship between the elasticity of growth $f_{\rm b}$, the proportion of maturing recruits to the mature age class $\gamma_y$ (which has a value of 0.50 in the simulated age-structured model), and the test-function $\lambda_1 \lambda_2$.
Our numerical results show that only Ricker-like SRRs can result in cyclic dynamics ($\lambda_1\lambda_2 \geq 1$; Fig. \ref{figest_age}d).
Moreover, we observe that cyclic dynamics can only emerge if $f_{\rm b} \leq -1$ for any potential value of $\gamma_y$, and this result applies to all potential SRRs.
Accordingly, as the ratio of maturing recruits declines (low $\gamma_y$; realized as the mortality of the mature age-class $M_y$ decreases), cyclic dynamics are less likely to occur unless the elasticity of growth is extremely low, which is biologically unreasonable.
As the ratio of maturing recruits increases (higher mature age-class mortality), the opposite occurs, and cyclic dynamics are more likely for a broader range of Ricker-like SRRs.
We have thus obtained a very powerful result: independent of the particular functions introduced into the general age-structured system, cyclic dynamics require 1) that spawning stock biomass includes a relatively large proportion of incoming recruits, and 2) that compensatory dynamics are driven by a Ricker-like function, where the elasticity of growth has a value $\leq -1$.
%By relating the elasticity of growth to stability regimes, knowledge of general aspects of the system - without assuming specific functional relationships - provides direct insights into the compensatory dynamics of age-structured populations.

%Comparison to real system
The relationship between recruitment, the elasticity of growth, and cyclic dynamics has predictive power, in particular because it is general without assumptions regarding the exact shapes of functional responses. 
For example, pink salmon ({\it Oncorhynchus gorbuscha}) are a widespread species with complex population dynamics \citep{Radchenko:2007tc}.
Depensatory dynamics are responsible for a large source of embryo mortality as spawning individuals compete for viable nests, such that Ricker-like models are generally predictive of stock recruitment relationships \citep{May:1974jz,Myers:1995gj}.
Moreover, pink salmon are semelparous, such that there is no overlap in spawning stock biomass between generations.
In our generalized modeling framework, this corresponds to an elasticity of growth $f_{\rm b} < 0$, and complete turnover of the adult age-class, such that $\gamma_y$ is close to unity.
From Fig. \ref{figest_age}d, we observe that cyclic dynamics are inevitable if $f_{\rm b} < -1$ as $\gamma_y \rightarrow 1$.
In nature, pink salmon populations are strongly cyclic, generally on the order of two-year cycles, and this is thought to be caused by density-dependent mortality reinforced by external sources of stochasticity \citep{Krkosek:2011ds}.
Thus, we observe that by relating the elasticity of growth to stability regimes, knowledge of general aspects of population dynamics - without assuming specific functional relationships - can provide direct insights into the compensatory dynamics of age-structured populations.

%A brief consideration of certain fisheries in nature suggests that this general relationship has some predictive power.
%For example, pink salmon are known to fluctuate regularly [details].
%Pink salmon are also known for having very low survivorship after the first year as an adult.
%This means that the mature age-class generally has a very high proportion of incoming recruits.
%In our generalized modeling framework, this correlates to high values of $\gamma_y$, and if the SRR is assumed to be Ricker-like, cyclic dynamics are a likely potential dynamic for the fisheries.
%Thus, we observe that by relating the elasticity of growth to stability regimes, knowledge of general aspects of the system - without assuming specific functional relationships - provides direct insights into the compensatory dynamics of age-structured populations.

\section{Discussion}

\begin{figure*}[h]
   \centering
   \includegraphics[width=0.65\textwidth]{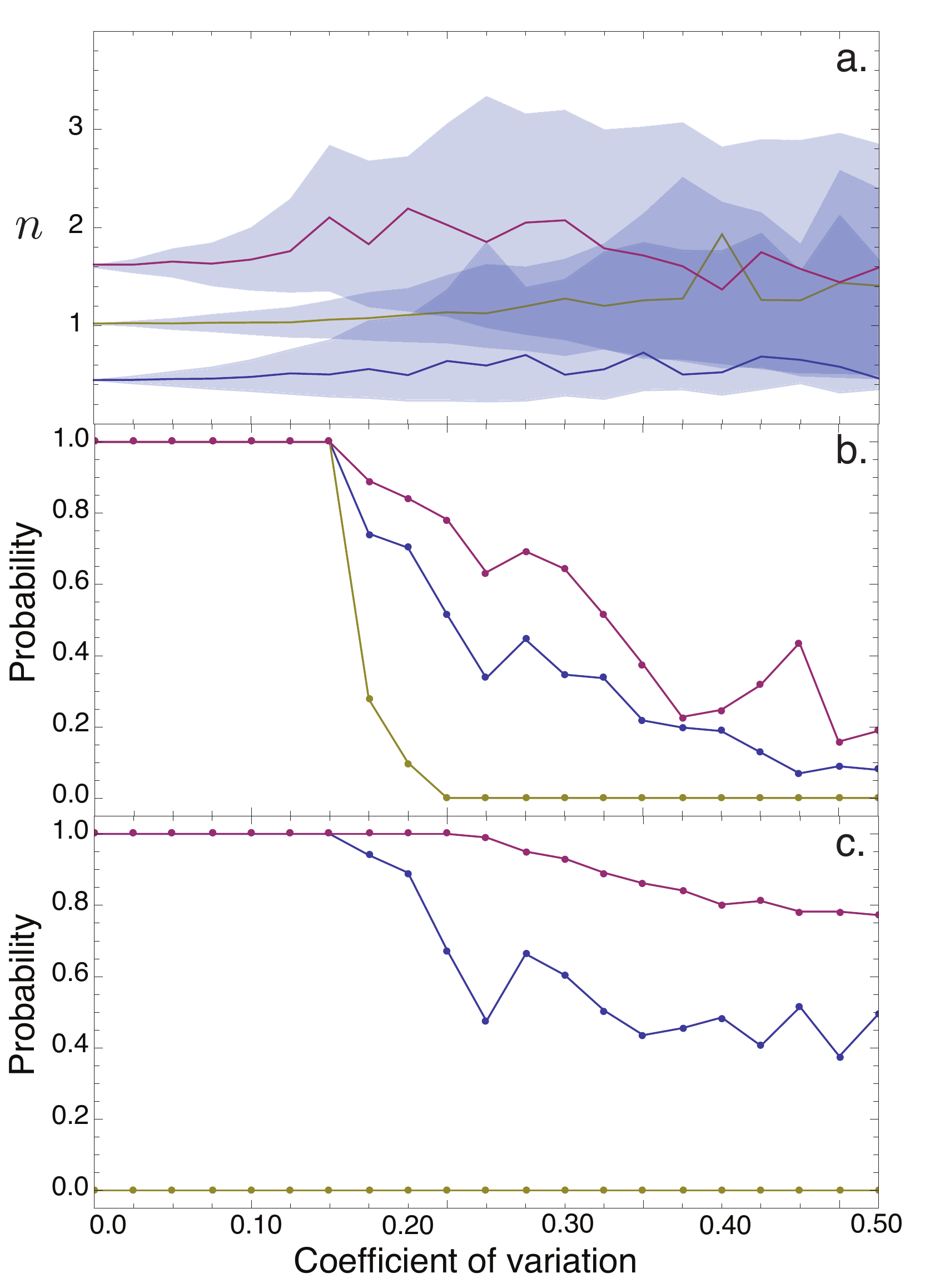}
      \caption{
      a. Estimation of the degree of compensation as a function of the coefficient of variation for 6300 simulated population trajectories (300 for each of 21 different values of CV).
      Shaded areas show the standard deviation of estimated $n$ values, while colored lines show the means (blue: Ricker; red: Cushing; yellow: B-H).
      b. The probability of correctly identifying the specific model from the others (Ricker: $n=0.5$; Cushing: $n=1.5$, and B-H: $n=1$).
      c. The probability of correctly identifying the SRR family (Ricker: $n<1$; Cushing: $n>1$; B-H: $n=1$).
      The probability of measuring the B-H SRR is always zero.
      }
      \label{figProb}
\end{figure*}

% What we have shown - summary
We have shown that the elasticity of growth in a generalized production model can be related directly to the degree of compensation parameter that determines Ricker-like, Cushing-like, or Beverton-Holt  behaviors.
%The elasticity of growth provides a nonlinear measure for the sensitivity of growth $s(b)$ to variations in biomass $b$, where both $s(b)$ and $b$ are normalized to the steady state $B^*$.
The elasticity of growth is useful because it is defined with respect to the biological and environmental conditions present during measurement, and thus can be estimated from limited time-series data.
Moreover, because large ranges of the elasticity of growth, and by extension the rate of relaxation, characterize families of functional forms, these measures are error tolerant (Fig \ref{figProb}c), particularly if the goal is to distinguish between SRRs with Ricker-like or Cushing-like recruitment dynamics.

The functional elasticities of both production and age-structured models can be used to determine directly the compensatory dynamics driving SRRs.
This method may be of most use to recent fisheries, where long-term time-series data do not yet exist.
Because we have employed elasticities in a generalized modeling framework, they are well-suited to inform knowledge of the general nature of compensation, and thus may be particularly useful for developing priors for parameters in flexible SRRs, such as the degree of compensation in the Shepherd model.
%Functional elasticities have certain qualities that aid interpretations of recruitment dynamics - in particular they do not require the formalization of recruitment across all biologically reasonable and unreasonable values of spawning stock biomass.
Determining SRRs from elasticities may also be useful if populations have highly variable recruitment dynamics, or dynamics that are strongly sensitive to changing environmental conditions, and it may be instructive to consider alternative approaches for measuring elasticities across a broader range of management scenarios.

\begin{acknowledgements}
We thank S. Allesina, M.P. Beakes D. Braun, T. Gross, C. Kuehn, T. Levi, A. MacCall, J.W. Moore, S. Munch, M. Novak, C.C. Phillis and A.O. Shelton for many helpful discussions and comments. We also thank the Dynamics of Biological Networks Lab at the Max-Planck Institute for the Physics of Complex Systems and the University of Bristol, for sharing the ideas and knowledge that inspired this work. This project was partially funded by the Center for Stock Assessment and Research, a partnership between the Fisheries Ecology Division, NOAA Fisheries, Santa Cruz, CA and the University of California, Santa Cruz and by NSF grant EF- 0924195 to M.M.
\end{acknowledgements}

\newpage

%\begin{figure}[h!]
%   \centering
%   \includegraphics[width=0.75\textwidth]{Figure_Functions.pdf}
%      \caption{
%       }
%      \label{figall}
%\end{figure}
%\newpage

\newpage

\newpage

\newpage

% s_b as the return to equilibrium
% sensitivity analysis

\newpage

{\renewcommand{\arraystretch}{2} %<- modify value to suit your needs

\begin{table}[h]
\caption{
Criteria for determining the elasticity of growth $s_{\rm b}$ from the rate of return to the steady state after a perturbation, $\lambda$, for the production model. The non-overlapping intervals for the elasticity of growth uniquely identify of Ricker-like, Beverton-Holt, and Cushing-like recruitment dynamics.
}
\begin{tabular}{l  l  l}
\hline\noalign{\smallskip}
Model & Elasticity ($B^*\gg0$) & Criterion \\
\noalign{\smallskip}\hline\noalign{\smallskip}
Ricker-like & $s_{\rm b} < 0~~{\rm s.t.}~~\frac{\lambda}{\gamma} + d_{\rm b} < 0$ & $\lambda < -\gamma d_{\rm b}$ \\
Beverton-Holt & $s_{\rm b} = 0~~{\rm s.t.}~~\frac{\lambda}{\gamma} + d_{\rm b} = 0$ & $\lambda = -\gamma d_{\rm b}$ \\
Cushing-like & $s_{\rm b} > 0~~{\rm s.t.}~~\frac{\lambda}{\gamma} + d_{\rm b} > 0$ & $\lambda > -\gamma d_{\rm b}$ \\
\noalign{\smallskip}\hline
\end{tabular}
\label{tab1}
\end{table}

\newpage
%\bibliographystyle{spbasic}
%\bibliography{shepherd}

\newpage


\begin{thebibliography}{35}
\providecommand{\natexlab}[1]{#1}
\providecommand{\url}[1]{{#1}}
\providecommand{\urlprefix}{URL }
\expandafter\ifx\csname urlstyle\endcsname\relax
  \providecommand{\doi}[1]{DOI~\discretionary{}{}{}#1}\else
  \providecommand{\doi}{DOI~\discretionary{}{}{}\begingroup
  \urlstyle{rm}\Url}\fi
\providecommand{\eprint}[2][]{\url{#2}}

\bibitem[{Beverton and Holt(1957)}]{Beverton:1957we}
Beverton RJH, Holt SJ (1957) {On the dynamics of exploited fish populations.}
  Fish and Fisheries, Vol. 11, Springer, New York

\bibitem[{Brooks and Powers(2007)}]{Brooks:2007cu}
Brooks EN, Powers JE (2007) {Generalized compensation in stock-recruit
  functions: properties and implications for management}. ICES J Mar Sci
  64(3):413--424

\bibitem[{Cushing(1973)}]{Cushing:1973wg}
Cushing DH (1973) {The dependence of recruitment on parent stock}. J Fish Res
  Bd Can 30:1965--1976

\bibitem[{Cushing(1988)}]{Cushing:1988vb}
Cushing DH (1988) {The problems of stock and recruitment}. In: Fish population
  dynamics: the implications for management, John Wiley {\&} Sons Inc

\bibitem[{Fell(1992)}]{Fell:1992uh}
Fell DA (1992) {Metabolic control analysis: a survey of its theoretical and
  experimental development.} Biochem J 286:313--330

\bibitem[{Fell and Sauro(1985)}]{FELL:1985p3430}
Fell DA, Sauro HM (1985) {Metabolic control and its analysis}. Eur J Biochem
  148(3):555--561

\bibitem[{Gross and Feudel(2006)}]{Gross:2006wn}
Gross T, Feudel U (2006) {Generalized models as a universal approach to the
  analysis of nonlinear dynamical systems.} Phys Rev E 73(1 Pt 2):016205

\bibitem[{Gross et~al(2009)Gross, Rudolf, Levin, and Dieckmann}]{Gross:2009jr}
Gross T, Rudolf L, Levin SA, Dieckmann U (2009) {Generalized models reveal
  stabilizing factors in food webs}. Science 325(5941):747--750

\bibitem[{Guckenheimer and Holmes(1983)}]{GuckHolmes}
Guckenheimer J, Holmes P (1983) {Nonlinear oscillations, dynamical systems, and
  bifurcations of vector fields}. Springer, Berlin, Heidelberg, New York

\bibitem[{Guill et~al(2011{\natexlab{a}})Guill, Drossel, Just, and
  Carmack}]{Guill:2011bf}
Guill C, Drossel B, Just W, Carmack E (2011{\natexlab{a}}) {A three-species
  model explaining cyclic dominance of Pacific salmon}. J Theor Biol
  276(1):16--21

\bibitem[{Guill et~al(2011{\natexlab{b}})Guill, Reichardt, Drossel, and
  Just}]{Guill:2011vx}
Guill C, Reichardt B, Drossel B, Just W (2011{\natexlab{b}}) {Coexisting
  patterns of population oscillations: the degenerate Neimark-Sacker
  bifurcation as a generic mechanism.} Phys Rev E 83(2):021910

\bibitem[{Gulland(1988)}]{Gulland:1988vb}
Gulland JA (1988) {The analysis of data and development of models}. In: Fish
  population dynamics: the implications for management, John Wiley {\&} Sons
  Inc

\bibitem[{Hilborn and Mangel(1997)}]{Hilborn:1997ds}
Hilborn R, Mangel M (1997) {The ecological detective: Confronting models with
  data}. Princeton University Press, Princeton

\bibitem[{Horvitz et~al(1997)Horvitz, Schemske, and Caswell}]{Horvitz:1997wy}
Horvitz C, Schemske DW, Caswell H (1997) {The relative ``importance'' of
  life-history stages to population growth: prospective and retrospective
  analyses}. In: Structured-population models in marine, terrestrial, and
  freshwater systems, Chapman and Hall, New York, pp 247--271

\bibitem[{Krkosek et~al(2011)Krkosek, Hilborn, Peterman, and
  Quinn}]{Krkosek:2011ds}
Krkosek M, Hilborn R, Peterman RM, Quinn TP (2011) {Cycles, stochasticity and
  density dependence in pink salmon population dynamics}. Proc Roy Soc B
  278(1714):2060--2068

\bibitem[{Kuehn et~al(2012)Kuehn, Siegmund, and Gross}]{Kuehn:2010p3380}
Kuehn C, Siegmund S, Gross T (2012) {Dynamical analysis of evolution equations
  in generalized models}. IMA J Appl Math

\bibitem[{Kuznetsov(1998)}]{AleksandrovichKuznetsov:1995p2580}
Kuznetsov Y (1998) {Elements of applied bifurcation theory}. Springer New York

\bibitem[{MacCall(2002)}]{MacCall:2002id}
MacCall AD (2002) {Use of known-biomass production models to determine
  productivity of west coast groundfish stocks}. N Am J Fish Manage
  22(1):272--279

\bibitem[{Mangel(2006)}]{Mangel:2006wa}
Mangel M (2006) {The theoretical biologist's toolbox: Quantitative methods for
  ecology and evolutionary biology}. Cambridge University Press, Cambridge

\bibitem[{Mangel et~al(2002)Mangel, Marinovic, Pomeroy, and
  Croll}]{Mangel:2002vh}
Mangel M, Marinovic B, Pomeroy C, Croll D (2002) {Requiem for Ricker: Unpacking
  MSY}. B Mar Sci 70(2):763--781

\bibitem[{Mangel et~al(2006)Mangel, Levin, and Patil}]{Mangel:2006tm}
Mangel M, Levin P, Patil A (2006) {Using life history and persistence criteria
  to prioritize habitats for management and conservation.} Ecol Appl
  16(2):797--806

\bibitem[{Mangel et~al(2010)Mangel, Brodziak, and DiNardo}]{Mangel:2010tp}
Mangel M, Brodziak J, DiNardo G (2010) {Reproductive ecology and scientific
  inference of steepness: a fundamental metric of population dynamics and
  strategic fisheries management}. Fish Fish 11:89--104

\bibitem[{Mangel et~al(2013)Mangel, MacCall, Brodziak, Dick, Forrest, Pourzand,
  and Ralston}]{Mangel:2013tu}
Mangel M, MacCall AD, Brodziak JK, Dick EJ, Forrest RE, Pourzand R, Ralston S
  (2013) {A perspective on steepness, reference points, and stock assessment}.
  Can J Fish Aquat Sci pp 1--64

\bibitem[{May(1974)}]{May:1974jz}
May RM (1974) {Biological populations with nonoverlapping generations: Stable
  points, stable cycles, and chaos}. Science 186(4164):645--647

\bibitem[{Morgan et~al(2011)Morgan, Perez-Rodriguez, Saborido-Rey, and
  Marshall}]{Morgan:2011eba}
Morgan MJ, Perez-Rodriguez A, Saborido-Rey F, Marshall CT (2011) {Does
  increased information about reproductive potential result in better
  prediction of recruitment?} Can J Fish Aquat Sci 68(8):1361--1368

\bibitem[{Munch et~al(2005)Munch, Kottas, and Mangel}]{Munch:2005dn}
Munch SB, Kottas A, Mangel M (2005) {Bayesian nonparametric analysis of
  stock-recruitment relationships}. Can J Fish Aquat Sci 62(8):1808--1821

\bibitem[{Myers et~al(1995)Myers, Barrowman, Hutchings, and
  Rosenberg}]{Myers:1995gj}
Myers RA, Barrowman NJ, Hutchings JA, Rosenberg AA (1995) {Population dynamics
  of exploited fish stocks at low population levels.} Science
  269(5227):1106--1108

\bibitem[{Radchenko et~al(2007)Radchenko, Temnykh, and
  Lapko}]{Radchenko:2007tc}
Radchenko VI, Temnykh OS, Lapko VV (2007) {Trends in abundance and biological
  characteristics of pink salmon ({\it Oncorhynchus gorbuscha}) in the North
  Pacific Ocean}. North Pac Anadromous Fish Comm Bull 4:7--21

\bibitem[{Ricker(1954)}]{Ricker:1954wpa}
Ricker W (1954) {Stock and recruitment}. Can J Fish Aquat Sci 11(5):559--623

\bibitem[{Shelton and Mangel(2011)}]{Shelton:2011eq}
Shelton AO, Mangel M (2011) {Fluctuations of fish populations and the
  magnifying effects of fishing.} Proc Natl Acad Sci USA 108(17):7075--7080

\bibitem[{Shepherd(1982)}]{Shepherd:1982p3428}
Shepherd J (1982) {A versatile new stock-recruitment relationship for
  fisheries, and the construction of sustainable yield curves}. J Conseil
  40(1):67--75

\bibitem[{Sissenwine and Shepherd(1987)}]{Sissenwine:1987td}
Sissenwine MP, Shepherd JG (1987) {An alternative perspective on recruitment
  overfishing and biological reference points}. Can J Fish Aquat Sci
  44(4):913--918

\bibitem[{Stiefs et~al(2010)Stiefs, van Voorn, Kooi, Feudel, and
  Gross}]{Stiefs:2010p2566}
Stiefs D, van Voorn GAK, Kooi BW, Feudel U, Gross T (2010) {Food quality in
  producer-grazer models: A generalized analysis}. Am Nat 176(3):367--380

\bibitem[{Sydsaeter and Hammond(1995)}]{Sydsaeter:2008vr}
Sydsaeter K, Hammond PJ (1995) {Essential Mathematics for Economic Analysis}.
  Prentice-Hall Inc., New Jersey

\bibitem[{Yeakel et~al(2011)Yeakel, Stiefs, Novak, and
  Gross}]{Yeakel:2011p3406}
Yeakel JD, Stiefs D, Novak M, Gross T (2011) {Generalized modeling of
  ecological population dynamics}. Theor Ecol 4(2):179--194

\end{thebibliography}

\end{document}